\documentclass[12pt]{article}
\usepackage{amsfonts,amsmath,amssymb,amscd,mathrsfs}

\usepackage{breakcites}

\usepackage[
backend=bibtex, 
natbib=true,
style=phys,
biblabel=brackets,
giveninits=true,
isbn=false,
block=space,
]{biblatex}

\AtEveryBibitem{\clearfield{month}}

\title{On the action principle as a guide to substantive general covariance}	
	
\author{Ward Struyve\footnote{Department of Physics and Astronomy, KU Leuven, Belgium}$^{*}$\footnote{Centre for Logic and Philosophy of Science, KU Leuven, Belgium}  }

\def\lam{\lambda}

\def\pa{\partial}

\newcommand{\be}{\begin{equation}}
\newcommand{\en}{\end{equation}}
\newcommand{\bi}{\begin{itemize}}
\newcommand{\ei}{\end{itemize}}

\begin{document}
\date{}

\maketitle

\begin{abstract}
\noindent
While Einstein was guided by the principle of general covariance in formulating general relativity, Kretschmann later argued that this principle lacks physical significance, since any space-time theory can be reformulated in a generally covariant form. This critique has prompted an ongoing debate over how to distinguish {\em substantive} general covariance from mere {\em formal} general covariance. Some proposals for defining substantive general covariance are based on the requirement that a theory be derivable from a diffeomorphism-invariant action. The present work aims to critically assess these proposals by examining canonical examples of Kretschmannian formulations of special relativistic theories. It will be shown that these formulations -- which seem merely formally generally covariant -- can always be derived from a diffeomorphism-invariant action. Although these actions involve auxiliary variables, these variables are trivial in the sense that they are either pure gauge or dynamically fixed by the other variables. Consequently, the theories described by these actions are arguably equivalent to the original Kretschmannian formulations. This stands in contrast to the actions proposed by Rosen and Sorkin, which include non-trivial auxiliary variables and hence seem to describe distinct physical theories. More recently, Freidel and Teh have suggested an additional criterion for substantial general covariance, namely that the action should also yield a non-trivial corner charge associated to diffeomorphism invariance. However, this too appears insufficient, since such actions can always be constructed.

\end{abstract}

\section{Introduction}
In developing general relativity, Einstein took the requirement of general covariance as a guiding principle. However, Kretschmann \cite{kretschmann17} soon objected that any space-time theory can be written in a generally covariant form, so that the requirement merely puts a constraint on the form and not the physical content of theories.{\footnote{See \cite{rynasiewicz99} for a in-depth discussion of Kretschmann's paper and  \cite{norton93} for a detailed review of the ensuing debate.}} As an example, consider a special relativistic theory involving tensor fields. The dynamical equations can be cast in generally covariant form by replacing the  Minkoskwi metric $\eta_{\mu \nu}$ by a Lorentzian metric $g_{\mu \nu}(x)$ on ${\mathbb R}^4$, substituting the space-time derivatives with covariant ones, and imposing the constraint that the Riemann curvature tensor vanishes, i.e.,
\be
R_{\mu \nu \rho \sigma} = 0.
\label{flatness}
\en
This constraint implies that the metric $g_{\mu \nu}$ is flat, i.e., it is given by the Minkowski metric up to a coordinate transformation. 

Kretschmann's critique has led to an ongoing debate over whether there exists a notion of {\em substantive general covariance}, which would be instantiated in general relativity, but not in Kretschmannian formulations \cite{anderson67,friedman83,norton93,earman06a,,earman06b,giulini07,pitts06,pooley10,pooley17,read24}. The latter would then be merely {\em formally generally covariant}. Some attempts to make the distinction rest on the question whether the theory can be derived from a diffeomorphism-invariant action (see \cite{pitts06,pooley17,read24} for excellent reviews). 
The goal of this paper is to critically asses these attempts.  In particular, it will be argued that Kretschmannian formulations of the type  above can always be derived from a diffeomorphism-invariant action. 

One of the earliest appeals to the action principle is by Anderson \cite{anderson67}, who advanced a prominent proposal for defining substantive general covariance. The basic idea is the following. Consider a theory that is diffeomorphism invariant, meaning that diffeomorphisms map solutions of the dynamical equations to solutions. For Anderson, this identifies the diffeomorphism group as the {\em covariance group}. This group is distinguished from the {\em invariance group}, which is the subgroup of the covariance group that leaves the absolute objects invariant. The absolute objects are the dynamical variables that are the same for any solution of the dynamical equations, up to a diffeomorphism. Hence, if there are no absolute objects, then the invariance group coincides with the diffeomorphism group and the theory can be regarded as substantially generally covariant. For Anderson, this was the distinctive feature of general relativity. In the Kretschmannian  example above, the metric is an absolute object, so that the symmetry group is not the group of diffeomorphisms, but the subgroup of Poincar\'e transformations. In other words, the theory is merely formally generally covariant. While Anderson's approach seems to establish an important distinction from general relativity, it is not without problems, since it seems to misclassify other theories, see Pitts \cite{pitts06} for an extensive discussion.{\footnote{Actually, adopting Friedman's refinement of Anderson's analysis \cite{friedman83}, even general relativity has an absolute object \cite{pitts06}.}}

Anderson \cite[p.\ 88]{anderson67} also claimed that the flatness condition \eqref{flatness} cannot be derived from an action principle (on the grounds that there are more equations of motion than independent components of the absolute object) and conjectured this to be a general feature of absolute objects \cite[p.\ 88]{anderson67}:
\begin{quote}
	The equations of motion for the dynamical objects can often be derived from a variational principle, especially if these objects are fields. On the other hand, it appears to be the case, although we can give no proof of the assertion, that the equations of motion for the absolute objects do not have this property.
\end{quote}
Anderson's claim about the flatness condition seems right, at least if no auxiliary variables are introduced. Such variables would obviously not be present in the equations of motion of interest and hence remain unobservable. They were labeled {\em irrelevant} by Anderson and excluded from the outset \cite[p.\ 83]{anderson67}. However, as will be discussed below, this exclusion seems problematic. While gauge variables clearly qualify as irrelevant, they seem indispensable to formulate an action for general relativity. 

So if Anderson's conjecture is correct, a theory that derives from an action principle does not contain absolute objects. This suggests that a theory will qualify as substantively generally covariant if it can be derived from an action that is diffeomorphism invariant. 

In recent years, the debate on substantive general covariance has sometimes shifted to the related debate on background independence (see \cite{pooley17,read24} for extensive reviews). If diffeomorphism invariance is not a distinctive feature of general relativity, then maybe it is the fact that it is background independent. But what exactly does it mean for a theory to be background independent? As with attempts to define substantive general covariance, various ideas have been explored. For example, Giulini suggests to take background independence as the absence of absolute objects in Anderson's sense~\cite{giulini07}. Another suggestion, discussed by Pooley, and again in the spirit of Anderson, is that \cite[p.\ 130]{pooley17}: 
\begin{quote}
	A theory T is background independent if
	and only if its solution space is determined by a generally covariant action: (i) all of whose dependent variables are subject to Hamilton's principle, and (ii) all of whose dependent variables represent physical fields.
\end{quote}
The demand that the action should not contain unphysical variables echoes Anderson's rejection of irrelevant variables. Indeed, Pooley's understanding of an unphysical variable appears to align with Anderson's idea of an irrelevant variable.

So, can the existence of an action principle serve as a criterion for substantive general covariance or background independence? Returning to the example of the Kretschmannian formulation above, the answer would seem to be negative if the flatness condition~\eqref{flatness} can be derived from an action principle. While this seems impossible without introducing auxiliary variables, Rosen 
\cite{rosen66} and Sorkin \cite{sorkin02} showed that it {\em can} be done if auxiliary variables are admitted (see again \cite{pitts06} for a detailed discussion). Rosen did this for the matter-free case and Sorkin for a scalar field. In both cases, the auxiliary field satisfies a non-trivial dynamics that depends on the metric, but there is no back-reaction on the metric or scalar field. The introduction of the auxiliary field implies that the solution space has been enlarged, even after identifying solutions related by diffeomorphisms. Hence, the corresponding theories appear to be different from the original Kretchmannian formulations determined by \eqref{flatness}, rather than mere reformulations \cite{earman06a,,earman06b,giulini07,rovelli07,pooley17}. Nevertheless, the theories by Rosen and Sorkin (with the auxiliary field) seem on par with the Kretschmannian formulation concerning the status of general covariance. Since the metric satisfies the same dynamics in both theories, they both seem to be merely formally and not substantively generally covariant \cite[p.\ 131]{pooley17}. So it seems that substantial general covariance is not guaranteed by merely requiring a diffeomorphism invariant action. However, the Rosen and Sorkin action could be dismissed on the basis that the auxiliary variable is irrelevant \cite{pitts06} or unphysical \cite{pooley17}. Namely, this variable does not affect the metric and scalar field and hence is unobservable if the empirical content resides in the latter.

The goal of this paper is to propose alternative actions. These still contain auxiliary variables, but unlike the Rosen and Sorkin actions, their dynamics is completely trivial: either these variables are completely unconstrained by the dynamical equations (derived as Euler--Lagrange equations from the action) and hence they can be regarded as gauge variables,{\footnote{This is according to the standard lore for singular Lagrangian systems, which yield a constrained Hamiltonian dynamics \cite{dirac64,hanson76,sundermeyer82,gitman90,henneaux92}.}} or the dynamical equations completely fix these variables in terms of the other variables. In both cases, the auxiliary variables can be seen as redundancies which merely correspond to different mathematical representations of the same physical situation. They are redundancies that can easily be eliminated on the level of the dynamical equations, but not on the level of the action, where they seem to be essential to allow for an action principle. 

There are various examples of theories that employ such redundant variables. First of all, Yang--Mills theories and general relativity contain gauge variables. In these cases, it is unclear how to eliminate them, even on the level of the dynamical equations, let alone on the level of the action. In general relativity, the diffeomorphisms are considered gauge transformations, where metrics connected by a diffeomorphism are regarded as physically equivalent (describing the same space-time geometry). The Einstein--Hilbert action of general relativity is formulated in terms of this gauge-dependent metric field. (In the ADM formalism, this becomes more explicit. The space-time metric is written in terms of the spatial metric and the lapse and shift functions. The dynamics of the latter is free and hence they are gauge variables. The spatial metric also contains gauge freedom, but it is harder to isolate.)

As an example of the second (non-gauge) type of redundancy, we can again consider the Einstein--Hilbert action. While this action is traditionally regarded as a functional of just the metric field, it can also be considered in the Palatini formulation as a functional of both the metric and the connection, treated as independent fields \cite{misner73,wald84}. The usual expression for the connection in terms of the metric is then obtained as an equation of motion. So this is an instance where the dynamics completely determines one of the variables, namely the connection, in terms of the other variables. But the Palatini formulation is not considered to yield a different physical theory. (Strictly speaking, there would still be an ontological difference if both the metric and the connection would be considered as part of the fundamental ontology, rather than just the metric. But that difference seems inessential.) Similarly, the electromagnetic action can be written in terms of the vector potential and the field strength as independent variables, with the usual expression for the field strength in terms of the vector potential arising as an equation of motion \cite{schwinger51,infeld54}. Another example is the reformulation of the scalar field theory in terms of a 5-component Kemmer spinor \cite{kemmer39,akhiezer65}. The Kemmer equation implies (in a particular representation of the Kemmer matrices) that one component is a scalar field that satisfies the Klein-Gordon equation, while the other components amount to the space-time derivatives of this scalar field. 

The auxiliary variables still qualify as irrelevant in Anderson's sense or as unphysical in Pooley's sense, but as just explained they are also present in our best physical theories like general relativity and Yang-Mills theories. It is not even clear how to formulate those theories without gauge variables (except in the Abelian case of the electromagnetic field). Therefore, while it seems reasonable to exclude actions with irrelevant or unphysical variables that are non-trivial, like the actions of Rosen and Sorkin, actions with trivial auxiliary variables should still be allowed. Otherwise there simply seems to be no hope of establishing substantive general covariance or background independence (through the existence of an action principle) even in the case of general relativity. 

In summary, auxiliary variables of the type just discussed can be regarded as mere representational redundancies. In particular, after identification of gauge equivalent solutions, the solution space is in a natural one-to-one correspondence with that of the original theory. As such, the theories described by the proposed actions are arguably mere reformulations of the original theory. The conclusion is that the requirement that the theory be derivable from a diffeomorphism-invariant action is not sufficient for substantive general covariance or background independence.  

Perhaps there are additional criteria that could be considered. Freidel and Teh have recently proposed one, namely that the action should also yield a non-trivial corner charge associated to diffeomorphism invariance \cite{freidel22} (discussed in more detail in \cite{teh24}). However, it will be shown that such actions can always be found for Kretschmannian theories of the type above. 

The outline of the paper is as follows. In the next section, it will be illustrated how actions can be formulated by introducing auxiliary variables whose dynamics is trivial, using the ideas presented in \cite{struyve23}. In sections~\ref{srosen} and~\ref{ssorkin}, this is applied to the Kretschmannian theories of vacuum Minkowski space-time and the scalar field on Minkowski space-time. In section~\ref{sfreidel}, the proposal of Freidel and Teh is considered. We conclude in section~\ref{conclusion}.

\section{Formulating an action}\label{action}
Formulating an action is straightforward when auxiliary variables are allowed \cite{struyve23}. The discrete case is considered here for simplicity, as the generalization to field theories is immediate.

Consider  dynamical equations of the form 
\be
f_i(q,\dot q, \ddot q, \dots ,t) = 0 ,\qquad i = 1,\dots M,
\label{1}
\en
with $q$ some configuration and the dots representing higher-order time derivatives of $q$. These equations may or may not be derivable from an action that is only a functional of $q(t)$. This is the inverse problem of Lagrangian mechanics. By introducing
 Lagrange multipliers $\lambda_i$ as auxiliary variables, they follow from the action
\be
S_1 = \int dt \lam_{i} f_i
\label{1.1}
\en
by varying with respect to $\lambda_i$. Variation with respect to $q$ yields the additional equations 
\be
\int dt \lam_i \frac{\delta f_i}{\delta q} =0.
\en
The variables $\lambda_i$ do not affect the $q_i$ and hence are unobservable if the empirical content of the theory resides in the $q_i$. Such variables would be deemed irrelevant by Anderson or unphysical by Pooley. In any case, their introduction entails that the theory defined by the action $S_1$ is different from the one given by the equations \eqref{1}. The Rosen and Sorkin actions are of this form, with Lagrange multipliers as extra variables that satisfy non-trival equations of motion. 

Another interesting (and early) application is the damped harmonic oscillator \cite{bateman31}. In that case the Lagrange multiplier satisfies the time-reversed of the damped harmonic oscillator equation. As such, as Morse and Feshbach note \cite[p.\ 298]{morse58}: ``the total energy is conserved, and we can have an invariant Lagrange function, at the sacrifice of a certain amount of ``reality" in some of the incidental results''. 

The Lagrange multipliers can also be introduced differently. Consider for example the action 
\be
S_2 = \int dt \lam_{ij} f_i f_j ,
\en
with $\lam_{ij} = \lam_{ji}$ as auxiliary variables. Variation with respect to the $\lam_{ij}$ yields the equations of motion \eqref{1}.
Variation with respect to $q$ does not yield additional  dynamical equations after taking into account $f_i=0$. So, the variables $\lam_{ij}$ remain free and as such they qualify as gauge variables. The dynamical equations admit the gauge symmetry 
\be
\lam_{ij}(t) \to \lam_{ij}(t) + \theta_{ij}(t),
\label{gauge}
\en
with $\theta_{ij}(t)$ arbitrary functions of time. After  identification of gauge-equivalent solutions, the solution space for this action is in a natural one-to-one correspondence with that of the dynamics \eqref{1}. So, the theory is arguably equivalent to the one determined by \eqref{1} (in contrast to the theory corresponding to $S_1$). 

Another possible action is
\be
S_3 = \int dt \left( \lam_{i} f_i +  \lam_{i} \lam_{j} {\bar \lam}_{ij} \right),
\en
with auxiliary variables $\lam_{i}$ and ${\bar \lam}_{ij} = {\bar \lam}_{ji}$. Variation with respect to $\lam_i$, ${\bar \lam}_{ij}$ and $q$ respectively yields
\be
f_i +2 \lam_j  {\bar \lam}_{ij} = 0, \qquad  \lam_i = 0, \qquad \int dt \lam_i \frac{\delta f_i}{\delta q} =0,
\en
and hence
\be
f_i = 0, \qquad \lam_i = 0.
\en
In this case, the auxiliary variables ${\bar \lam}_{ij}$ are gauge, while the $\lam_i$ are zero because of the dynamical equations. Again, identification of gauge equivalent solutions yields the original theory.

The form of the actions $S_2$ and $S_3$ also appears in the context of generalized $\xi$-gauges in the path integral approach to quantum field theory, see e.g.~\cite[p.\ 23]{weinberg96a}. Namely, if $f_i = 0$ is regarded as a gauge fixing condition for some theory described by an action $S$, then a term of the form  $S_2$ with $\lam_{ij}$ fixed parameters can be added to break the gauge symmetry. Alternatively, a term of the form $S_3$ can be added, with $\lam_i$ dynamical and  ${\bar \lam}_{ij}$ fixed parameters. The $\lam_i$ then play the role of Nakanishi-Lautrup fields. Our approach is different in that we take all the variables as dynamical, rather than fixed parameters. 

Finally, note that while the actions $S_2$ and $S_3$ yield dynamical equations with a gauge symmetry, the actions themselves are not invariant under the gauge transformations. For example, the dynamical equations corresponding to $S_2$ 
are invariant under the transformations \eqref{gauge}, but the action is not. However, it might be desirable to have that gauge symmetries of the dynamical equations correspond to variational symmetries of the action. This might perhaps serve as an additional requirement for substantial diffeomorphism invariance. This will not be further investigated here, but just as a toy example of how this could be achieved, consider the action $S=  \int dt \exp(\lam) q^2$. The corresponding Euler-Lagrange equation is $q=0$, while $\lam$ is free. This action is not invariant under the gauge transformation $\lam \to \lam + \epsilon$, but taken together with $q \to q - q \epsilon/2$, it is. On-shell (i.e., using the equations of motion) this reduces to just $\lam \to \lam + \epsilon$, as desired. So, at least in this case, there is a variational symmetry that corresponds to the gauge symmetry of the dynamical equations. It is unclear whether this can always be achieved.

\section{Action for Minkowski space-time}\label{srosen}
Consider the Kretschmannian description of Minkowski space-time in terms of a Lorent\-zi\-an metric $g_{\mu \nu}$ (with ``mostly minus'' convention) satisfying the flatness condition \eqref{flatness}. Rosen showed that \eqref{flatness} can be derived from the action \cite{rosen66}: 
\be
S_{\textrm{Rosen}} = \frac{1}{4}\int d^4x \sqrt{-g} \lambda^{\mu \nu \rho \sigma} R_{\mu \nu \rho \sigma},
\en
where $\lambda^{\mu \nu \rho \sigma}$ is an auxiliary field with the same symmetries as the curvature tensor, i.e., $\lambda^{\mu \nu \rho \sigma} = \lambda^{  \rho \sigma   \mu \nu} = - \lambda^{\nu \mu \rho \sigma} $ and $\lambda^{\mu \nu \rho \sigma} + \lambda^{\mu \rho \sigma \nu } + \lambda^{\mu  \sigma \nu \rho} =0$. Varying the action with respect to $\lam^{\mu \nu \rho \sigma}$ yields \eqref{flatness}, while variation with respect to $g_{\mu \nu}$ and using \eqref{flatness} yields
\be
\nabla_\rho \nabla_\sigma \lambda^{\mu \rho \nu  \sigma} =0.
\label{75}
\en
Hence, $\lambda^{\mu \nu \rho \sigma}$ satisfies a non-trivial dynamics, which implies that the theory actually differs from the Kretschmannian formulation in terms of just the flatness condition. As noted by Rosen, particular solutions of \eqref{75} are given by
\be
 {\widetilde \lambda}_c^{\mu \rho \nu  \sigma} = c (g^{\mu \sigma} g^{\rho \nu} - g^{\mu \nu} g^{\rho \sigma}),
 \label{con}
\en
with $c$ constant. 

As explained in the preceding section, a different action is
\be
S_{4} = \int d^4x \sqrt{-g} \lambda^{\mu \nu \rho \sigma \alpha \beta \gamma \delta} R_{\mu \nu \rho \sigma} R_{\alpha \beta \gamma \delta},
\en
with $\lambda^{\mu \nu \rho \sigma \alpha \beta \gamma \delta}$ the same symmetries as $R^{\mu \nu \rho \sigma} R^{\alpha \beta \gamma \delta}$. The corresponding dynamics is just the flatness condition \eqref{flatness}, while the auxiliary field is free and hence a gauge variable. 

Another possible action is
\be
S_{5} = \int d^4x \sqrt{-g}\left[\frac{1}{4} \lambda^{\mu \nu \rho \sigma} R_{\mu \nu \rho\sigma} + \left(\lambda^{\mu \nu \rho \sigma} - {\widetilde \lambda}_c^{\mu \nu \rho \sigma}\right) \left(\lambda^{\alpha \beta \gamma \delta} - {\widetilde \lambda}_c^{\alpha \beta \gamma \delta}\right) {\bar \lambda}_{\mu \nu \rho \sigma \alpha \beta \gamma \delta}\right] , 
\en
where ${\widetilde \lambda}_c^{\mu \rho \nu  \sigma}$ is the metric-dependent expression given in \eqref{con} and (as before) the auxiliary fields have the appropriate symmetry properties to avoid redundancies. Variation with respect to ${\bar \lambda}_{\mu \nu \rho \sigma \alpha \beta \gamma \delta}$ yields 
\be
\lambda^{\mu \nu \rho \sigma} = {\widetilde \lambda}_c^{\mu \nu \rho \sigma}.
\label{100}
\en 
Variation with respect to $\lambda^{\mu \nu \rho \sigma}$ and using \eqref{100} yields \eqref{flatness}. Finally, variation with respect to $g_{\mu \nu}$ does not yield an extra equation given \eqref{flatness} and \eqref{100}. Summarizing, the desired flatness condition is obtained, while the field $\lambda^{\mu \nu \rho \sigma}$ is a fixed function of the metric, and the field ${\bar \lambda}_{\mu \nu \rho \sigma \alpha \beta \gamma \delta}$ is gauge. A particularly simple choice is of course to take $c=0$, so that ${\widetilde \lambda}_0^{\mu \rho \nu  \sigma} =0$ and \eqref{100} amounts to $\lambda^{\mu \nu \rho \sigma} = 0$.

As explained in the previous section, the theories corresponding to the actions $S_4$ and $S_5$ amount to the Kretchmannian theory given by the flatness condition.

\section{Scalar field on Min\-kow\-ski space-time}\label{ssorkin}
Sorkin's action \cite{sorkin02} is obtained by adding the scalar field action $S_M$ to Rosen's action:
\be
S_{\textrm{Sorkin}} =  S_M + S_{\textrm{Rosen}}= \int d^4 x \sqrt{-g} \left(-\frac{1}{2} g^{\mu \nu} \nabla_\mu \phi \nabla_\nu \phi  + \frac{1}{4}  \lambda^{\mu \nu \rho \sigma} R_{\mu \nu \rho \sigma} \right).\label{s0}
\en
The corresponding  dynamical equations are
\be
g^{\mu \nu} \nabla_\mu \nabla_\nu \phi = 0 ,\qquad R_{\mu \nu \kappa \sigma} = 0,
\label{s1}
\en
\be
 \nabla_\rho \nabla_\sigma \lambda^{\mu  \rho \nu \sigma} = T^{\mu \nu },
\label{s2}
\en
with $ T^{\mu \nu} = \nabla^\mu \phi \nabla^\nu \phi - g^{\mu \nu} \nabla_\rho \phi \nabla^\rho \phi/2  $ the energy-momentum tensor of the scalar field. Because of the flatness condition, these equations amount to the Klein--Gordon equation on flat space-time, together with an equation for the auxiliary field $ \lambda^{\mu \nu \rho \sigma}$. 

Sorkin demonstrated that \eqref{s2} does not put further restrictions on the scalar field,  by proving the existence of solutions $\lambda^{\mu  \rho \nu \sigma}$ for a given energy-momentum tensor. A particular solution can be given by considering a coordinate system where the metric $g_{\mu \nu}$ equals the Minkowski metric $\eta_{\mu \nu }$. In such a coordinate system, a solution is
\be
{\widetilde \lambda}^{\mu \rho \nu  \sigma} = \eta^{\rho \sigma} \frac{1}{\square} T^{\mu \nu } + \eta^{\mu \nu } \frac{1}{\square} T^{\rho \sigma} - \eta^{\mu \sigma} \frac{1}{\square} T^{\rho \nu} - \eta^{\rho \nu} \frac{1}{\square} T^{\mu \sigma}  
\label{250},
\en
where
\be
\frac{1}{\square} f(x) = -\frac{1}{4 \pi} \int d^4 y \delta\big(\eta_{\mu \nu} (x^\mu-y^\mu)(x^\nu-y^\nu) \big) f(y)\label{250.1}
\en
is the convolution of $f$ with the time-symmetric Green function of the d'Alembertian. Namely, $\pa_\rho \pa_\sigma {\widetilde \lambda}^{\mu \rho \nu  \sigma} = \pa_\rho \pa_\sigma  \left( \eta^{\rho \sigma} \frac{1}{\square} T^{\mu \nu }\right) = T^{\mu \nu }$, where the first equality is obtained using the conservation of the energy-momentum tensor. To express ${\widetilde \lambda}^{\mu \rho \nu  \sigma} $ in a general coordinate system, the metric $\eta_{\mu \nu}$ should be replaced by $g_{\mu \nu}$ in \eqref{250} and \eqref{250.1}. To any particular solution one can of course always add an expression of the form ${\widetilde \lambda}_c^{\mu \rho \nu  \sigma}$ given in \eqref{con}. The field ${\widetilde \lambda}^{\mu \rho \nu  \sigma}$ is a non-local functional of $\phi$ and $g_{\mu \nu }$. Presumably, there is no local functional that solves \eqref{s2}. This is in contrast to the solutions ${\widetilde \lambda}_c^{\mu \rho \nu  \sigma}$ of \eqref{75} which are local functionals.

We now turn to actions with auxiliary variables whose dynamics is trivial, so that the corresponding theories just amount to one of a scalar field on Minkowski space-time described by the dynamical equations~\eqref{s1}. The action
\be
S_6 =  \int d^4 x \sqrt{-g} \left[ \lambda\left( g^{\mu \nu} \nabla_\mu \nabla_\nu \phi\right)^2  +  {\bar \lambda}^{\mu \nu \rho \sigma \alpha \beta \gamma \delta} R_{\mu \nu \rho \sigma} R_{\alpha \beta \gamma \delta} \right] 
\en
yields the desired equations of motion \eqref{s1} and leaves $\lambda$ and ${\bar \lambda}^{\mu \nu \rho \sigma \alpha \beta \gamma \delta}$ as gauge. 

Another possibility is 
\begin{multline}
S_7 =  \int d^4 x \sqrt{-g} \Bigg[ -\frac{1}{2} g^{\mu \nu} \nabla_\mu \phi \nabla_\nu \phi  + \frac{1}{4}  \lambda^{\mu \nu \rho \sigma} R_{\mu \nu \rho \sigma}  \\
+  \left(\lambda^{\mu \nu \rho \sigma} - {\widetilde \lambda}^{\mu \nu \rho \sigma}\right) \left(\lambda^{\alpha \beta \gamma \delta} - {\widetilde \lambda}^{\alpha \beta \gamma \delta}\right) {\bar \lambda}_{\mu \nu \rho \sigma \alpha \beta \gamma \delta}\Bigg],
\end{multline}
where ${\widetilde \lambda}^{\mu \nu \rho \sigma}$ is the solution of \eqref{s2} given in \eqref{250}. The presence of ${\widetilde \lambda}^{\mu \nu \rho \sigma}$ in the action implies that the action is a non-local one (i.e., the corresponding Lagrangian density is not a local functional of the fields), unlike the previous actions (or the Einstein--Hilbert action). Variation of $S_7$ with respect to $\phi$ yields the Klein--Gordon equation. Variation with respect to ${\bar \lambda}_{\mu \nu \rho \sigma \alpha \beta \gamma \delta}$ yields
\be
\lambda^{\mu \nu \rho \sigma} = {\widetilde \lambda}^{\mu \nu \rho \sigma}.
\label{150}
\en
Variation with respect to $\lambda^{\mu \nu \rho \sigma}$ and using \eqref{150} yields the flatness condition. Finally, variation with respect to $g_{\mu \nu}$, together with the flatness condition and \eqref{150}, yields
\be
\nabla_\rho \nabla_\sigma {\widetilde  \lambda}^{\mu  \rho \nu \sigma} = T^{\mu \nu},
\en
which is automatically satisfied by virtue of the definition of ${\widetilde \lambda}^{\mu \nu \rho \sigma}$. Summarizing, again \eqref{s1} is obtained, ${\bar \lambda}_{\mu \nu \rho \sigma \alpha \beta \gamma \delta}$ is gauge and $\lambda^{\mu \nu \rho \sigma}$ is a fixed function of the metric and scalar field.

Finally, for the purposes of the next section, consider the action
\begin{multline}
	S_8 =  \int d^4 x \sqrt{-g}  \lambda\left( g^{\mu \nu} \nabla_\mu \nabla_\nu \phi\right)^2   + S_5 
	=  \int d^4 x \sqrt{-g}  \Bigg[ \lambda\left( g^{\mu \nu} \nabla_\mu \nabla_\nu \phi\right)^2 \\
	+\frac{1}{4} \lambda^{\mu \nu \rho \sigma} R_{\mu \nu \rho\sigma} + \left(\lambda^{\mu \nu \rho \sigma} - {\widetilde \lambda}_c^{\mu \nu \rho \sigma}\right) \left(\lambda^{\alpha \beta \gamma \delta} - {\widetilde \lambda}_c^{\alpha \beta \gamma \delta}\right) {\bar \lambda}_{\mu \nu \rho \sigma \alpha \beta \gamma \delta}\Bigg].
\end{multline}
This is again a local action. The corresponding equations of motion are the Klein-Gordon equation and the flatness condition, $\lambda^{\mu \nu \rho \sigma} = {\widetilde \lambda}_c^{\mu \nu \rho \sigma}$, and $\lambda$ and ${\bar \lambda}_{\mu \nu \rho \sigma \alpha \beta \gamma \delta}$ are gauge fields.

Sorkin remarked that his action has the desirable feature that the conservation of $T^{\mu \nu}$ follows from diffeomorphism invariance of the action through a Noetherian analysis. This can be derived from the property that the matter part $S_M$ is separately diffeomorphism invariant, together with the fact that under a variation of the metric:
\be
\delta S_M = \frac{1}{2} \int d^4 x \sqrt{-g} T^{\mu \nu}\delta g_{\mu \nu},
\label{em}
\en
see e.g.\ \cite[p.\ 56]{carlip19}. While all the alternative actions that we considered are diffeomorphism invariant, it is only for the action $S_7$ that $T^{\mu \nu}$ satisfies \eqref{em} and that its conservation follows similarly from diffeomorphism invariance.

In this section, we considered the scalar field, but it is clear that other special relativistic theories can be similarly Kretschmannized. Theories involving tensorial fields, for instance, can also be reformulated in terms of a general curved metric, allowing for actions analogous to those discussed here. These techniques can even be applied to non-relativistic theories. For example, for the heat equation this can be done by first reformulating it in special relativistic form, by introducing a vector field $n^\mu$ that satisfies $\eta_{\mu \nu} n^\mu n^\nu =1$ and $\pa_\nu n^\mu =0$ \cite[p.\ 79]{anderson67}.

\section{Non-trivial corner charge as a supplementary criterion}\label{sfreidel}
In the preceding sections, it was shown that having a diffeomorphism-invariant action appears insufficient to classify a theory as substantively generally covariant. Perhaps supplementary criteria could be introduced. Recently Freidel and Teh argued that the action should also imply a non-trivial corner charge with respect to space-time diffeomorphisms \cite{freidel22}. 

To see how a corner charge may arise, let us summarize the main steps. Consider an action $S[\varphi]=\int d^4 x {\mathcal L}(\varphi(x),\partial_\mu \varphi (x), \dots)$ which is defined in terms of a Lagrangian density ${\mathcal L}$ which is a local functional of some fields which are collectively denoted as $\varphi$. Under a variation $\delta \varphi$, we have 
\be
\delta {\mathcal L} = - E_\varphi \delta \varphi + \pa_\mu \theta^\mu,
\en
where $E_\varphi$ denote the Euler-Lagrange expressions and $\theta^\mu$ is the symplectic potential current which depends on $\varphi$, $\delta \varphi$, and their derivatives,. If the action is invariant under infinitesimal diffeomorphisms $x^\mu \to x^\mu + \xi^\mu$, $\varphi \to \delta_\xi \varphi$, then Noether's second theorem implies the conserved current
\be
j^\mu_\xi  = \theta^\mu_\xi - l^\mu_\xi,
\en
where $\theta^\mu_\xi$ is obtained from $\theta^\mu$ by replacing $\delta \varphi$ by $\delta_\xi \varphi$, and $l^\mu_\xi$ is determined by $\delta_\xi {\mathcal L} = \pa_\mu l^\mu_\xi$. This current can be written as 
\be
j^\mu_\xi  = C^\mu_\xi + \pa_\nu U^{\mu \nu}_\xi,
\en
where $C^\mu_\xi$ is zero on-shell, and $U^{\mu \nu}_\xi = - U^{\nu \mu }_\xi $, so that $\pa_\nu U^{\mu \nu}_\xi$ is trivially conserved, irrespective of the dynamical equations. The tensor $U^{\mu \nu}_\xi$ is called the superpotential. A non-trivial superpotential implies a corner charge $\int_\Sigma d\sigma_\mu  \pa_\nu U^{\mu \nu}_\xi= \int_{\partial \Sigma} d\sigma_{\mu\nu}  U^{\mu \nu}_\xi$, where $\Sigma$ is a Cauchy hypersurface.

In the case of Sorkin's action \eqref{s0}, the symplectic potential current is
\be
\theta^\mu = - \sqrt{-g} \nabla^\mu \phi \delta \phi + \frac{\sqrt{-g}}{2} \left( \lambda^{\mu \alpha \beta \sigma} \nabla_\sigma \delta g_{\alpha \beta} - \nabla_\sigma \lambda^{\mu \alpha \beta \sigma}  \delta g_{\alpha \beta} \right). 
\en
Hence, with $\delta_\xi \phi = \pa_\mu \phi \xi^\mu$, $\delta_\xi g_{\mu \nu} = \nabla_\mu \xi_\nu + \nabla_\nu \xi_\mu$, and $l^\mu_\xi= -{\mathcal L}_{\textrm{Sorkin}} \xi^\mu$, the conserved current is
\be
j^\mu_\xi  = \sqrt{-g} \left[ - T^\mu_{\ \, \,\nu} \xi^\nu -  \nabla_\sigma \lambda^{\mu (\alpha \beta) \sigma}\nabla_\alpha \xi_\beta + \lambda^{\mu (\alpha \beta) \sigma} \nabla_\sigma \nabla_\alpha \xi_\beta - \frac{1}{4} \lambda^{\mu \nu \rho \sigma} R_{\mu \nu \rho \sigma} \xi^\mu\right].
\en
With some algebra, this leads to the superpotential
\be
U^{\mu \nu}_\xi= \sqrt{-g} \left(\nabla_\rho \lambda^{\mu \nu \rho \sigma} \xi_\sigma - \frac{1}{2} \lambda^{\mu \nu \rho \sigma} \nabla_\rho \xi_\sigma \right).
\label{500}
\en
Since this superpotential is non-trivial, there is a corresponding corner charge, and Freidel and Teh conclude that this theory is substantively generally covariant.

This conclusion seems counterintuitive. Sorkin's theory tends to be regarded as an example of a theory that is {\em not} substantively generally covariant \cite{pooley17}. Moreover, without introducing auxiliary variables, the flatness condition is not derivable from an action and as such the theory does not qualify as generally covariant for Freidel and Teh. It is only {\em with} the auxiliary variable  $\lambda^{\mu \nu \rho \sigma}$ that Sorkin's theory can be regarded as substantively generally covariant, even though the introduction of the auxiliary variable does not change the dynamical equations of the metric and scalar field. Freidel and Teh acknowlegde this, but argue that the introduction of $\lambda^{\mu \nu \rho \sigma}$ makes it a different theory, justifying a different classification.

However, this stance also means that Kretschmann's critique regarding the triviality of general covariance looms again. While certain Kretschmannian theories might not qualify as substantively generally covariant according to Freidel and Teh, there might be empirically equivalent theories that are. Rosen and Sorkin have actually provided us with a recipe to do so for a broad class of special relativistic theories. (It will not be applicable to {\em all} special relativistic theories. For example, stochastic theories simply do not admit an action principle. Other theories do not admit a {\em local} action and hence do not allow for a Noetherian analysis.)  

Moreover, as we saw in the preceding section, Kretschmannian theories can also be derived from actions that merely involve trivial auxiliary variables. Three examples were given: $S_6$, $S_7$ and $S_8$. The action $S_6$ implies a trivial corner charge. This immediately follows from the fact that the action is quadratic in the field equations and hence the conserved Noether current $j^\mu_\xi $ is proportial to the field equations, so that is vanishes on-shell, implying a vanishing superpotential. The action $S_7$ is non-local and hence does not allow for the standard Noetherian analysis. Finally, the action $S_8$ is a local one and implies again the superpotential \eqref{500}. (It is actually only the part $S_5$ that contributes to it.) On-shell, using the equation  $\lambda^{\mu \nu \rho \sigma} = {\widetilde \lambda}_c^{\mu \nu \rho \sigma}$, the superpotential reads
\be
U^{\mu \nu}_\xi= \sqrt{-g}\frac{c}{2} \left( \nabla^\mu \xi^\nu - \nabla^\nu \xi^\mu  \right).
\en
If $c$ is zero, the superpotential vanishes and there is no (non-trivial) corner charge. But if $c$ is non-zero, there is a non-trivial corner charge and the theory qualifies as substantively generally covariant according to Freidel and Teh. This result is quite general, since an action of the form $S_8$ can be constructed for a broad class of special relativistic theories. Hence, even requiring a non-trivial corner charge appears insufficient for substantial general covariance.

\section{Conclusion}\label{conclusion}
General relativity can be derived from an action that is both diffeomorphism-invariant and has a non-trivial corner charge associated to the diffeomorphism invariance. It has been proposed that these features capture the defining criteria of substantive general covariance. However, as was argued here, this does not appear to be the case. Actions with these properties can be formulated for theories that do not seem to qualify as substantively generally covariant. Although these actions require the introduction of auxiliary variables, these variables can be regarded as mere descriptive redundancies -- the type of redunancies that already seems indispensable in the formulation of general relativity itself.

\section{Acknowledgments}
It is a pleasure to thank Harvey Brown, Henrique Gomes and James Read for stimulating discussions. This work is supported by the Research Foundation Flanders (Fonds Wetenschappelijk Onderzoek, FWO), Grant No.\ G0C3322N. 

\printbibliography
\end{document}